%% file: main.tex
\documentclass[conference]{IEEEtran}
\IEEEoverridecommandlockouts
\usepackage{booktabs}

\usepackage{cite}

\usepackage{amsmath,amssymb,amsfonts}
\usepackage{algorithmic}
\usepackage{graphicx}
\usepackage{textcomp}
\usepackage{xcolor}
\usepackage{hyperref}
\usepackage{float}
\usepackage{dblfloatfix}
\usepackage{makecell}
\usepackage[normalem]{ulem}
\usepackage{soul}
\usepackage{amsmath}
\usepackage{tabularx}
\usepackage{svg}

\usepackage{multirow}
\usepackage{enumitem, array}
\usepackage{caption}
\usepackage{subcaption}
\usepackage[export]{adjustbox}
\usepackage{hyperref}
\usepackage{gensymb}
\usepackage{todonotes}
\usepackage{subfiles}

\graphicspath{{images/}}

\def\xhat{{ \mathbf{\hat{x}} }}
\def\xe{{ \mathbf{x}_e }}

\def\fsim{{ f_{\mathrm{sim}} }}
\def\fpcorr{{ f_{\mathrm{pcorr}} }}
\def\fssim{{ f_{\mathrm{ssim}} }}

\def\f{{\bf f}}
\def\r{{\bf r}}
\def\x{{\bf x}}

\def\y{{\bf y}}
\def\t{{\bf t}}

\def\loss{{\mathcal L}}

\def\BibTeX{{\rm B\kern-.05em{\sc i\kern-.025em b}\kern-.08em
    T\kern-.1667em\lower.7ex\hbox{E}\kern-.125emX}}

\begin{document}

\title{
    Assessing the Viability of Synthetic Physical Copy Detection Patterns on Different Imaging Systems
    \thanks{S. Voloshynovskiy is a corresponding author.}
}

\author{\IEEEauthorblockN{Roman Chaban, Brian Pulfer and Slava Voloshynovskiy}
\IEEEauthorblockA{Department of Computer Science, University of Geneva, Switzerland \\
\{roman.chaban, brian.pulfer, svolos\}@unige.ch}
}

\maketitle

\begin{abstract}
This paper explores the potential of synthetic physical Copy Detection Patterns (CDP) to improve the robustness of anti-counterfeiting systems. By leveraging synthetic physical CDP $\xhat$, we aim at enhancing security and cost-effectiveness across various real-world applications. Our research demonstrates that synthetic CDP offer substantial improvements in authentication accuracy compared to one based on traditional digital templates $\t$. We conducted extensive tests using both a scanner and a diverse range of mobile phones, validating our approach through ROC analysis. The results indicate that synthetic CDP can reliably differentiate between original and fake samples, making this approach a viable solution for real-world applications, though requires an additional research to make this technology scalable across a variety of imaging devices.
\end{abstract}


\subfile{sections/01_introduction}
\subfile{sections/02_dataset}
\subfile{sections/03_methodology}
\subfile{sections/04_results}
\subfile{sections/05_conclusions}

\bibliographystyle{IEEEtran}

\bibliography{IEEEexample}

\end{document}

%% file: sections/01_introduction.tex
\section{Introduction}

Copy Detection Patterns (CDP) have emerged as a promising anti-counterfeiting technology due to their cost-effectiveness, scalability, and ease of integration across various industries \cite{picard2004,picard2008copy}. These patterns are applicable to packaging, security labels, pharmaceuticals, currency, and identification documents, with their effectiveness stemming from their resistance to replication, machine verifiability, and versatility.

However, the sophisticated machine learning attacks have proven to be highly efficient against existing CDP, compromising their security \cite{chaban2021fakes,taran2019clones,tkachenko2019estimation,khermaza2021can}. These attacks utilize advanced algorithms to create high-quality fake CDP $\f$ that can bypass existing authentication methods, raising concerns about CDP's non-cloneability, particularly in critical sectors like pharmaceuticals. Meanwhile, authentication based on reference CDP $\xe$, a.k.a. physical template, offers superior protection in contrast to one based on digital template $\t$ \cite{chaban2024eusipco} but requires the enrollment of $\xe$, which compromises cost-effectiveness and should be used only in very specific applications.

\begin{figure}[t!]
    \centering
    \includegraphics[width=0.45\textwidth]{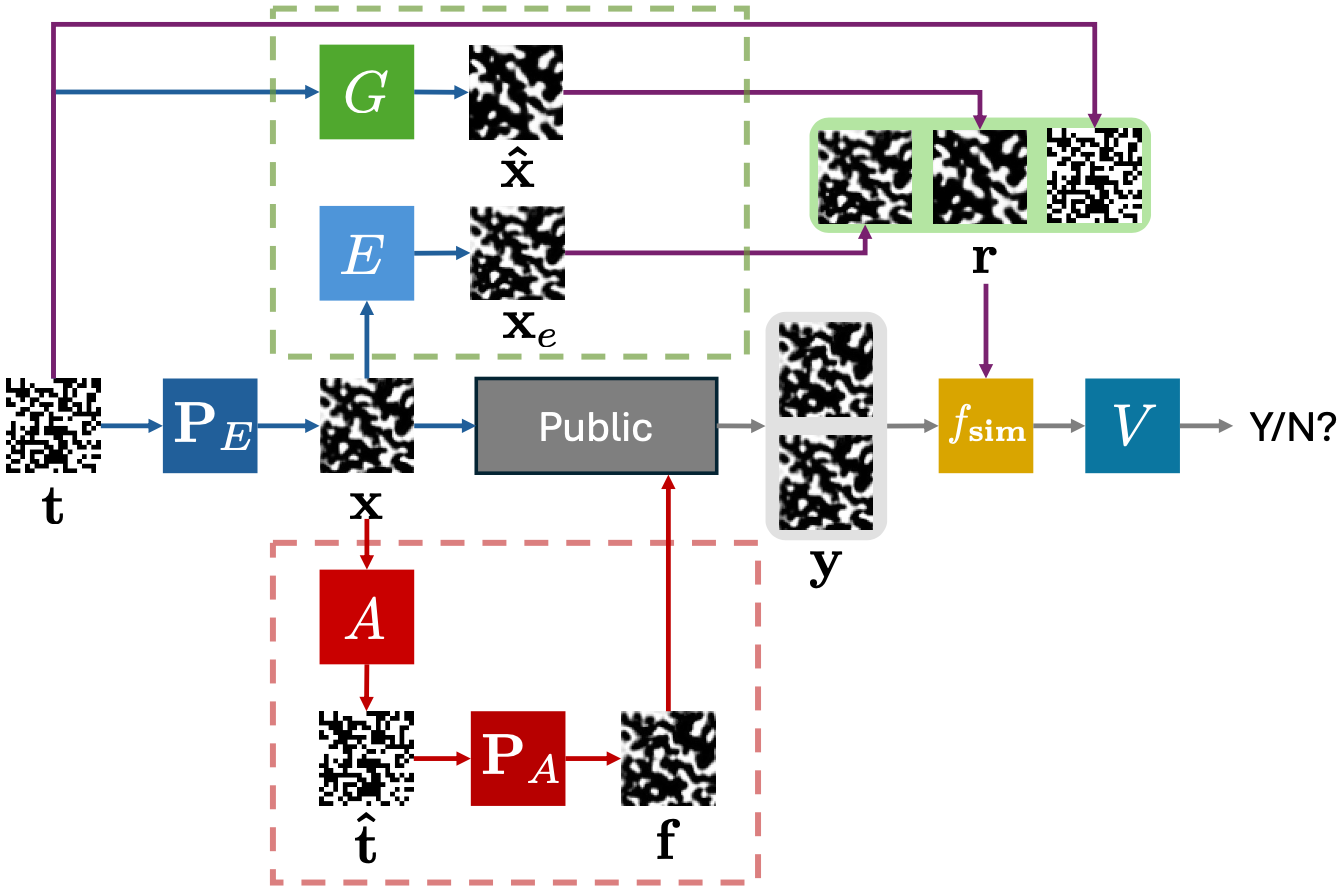}
    \caption{The diagram demonstrating the real-world CDP lifecycle which includes the authentication of probe $\y$, the possible attack scenario (red dashed line) yielding fake $\f$, and the proposed defensive strategies (green dashed line) based on the reference $\r \in \{\t, \xe, \xhat\}$.}
    \label{fig:lifecycle}
    \vspace{-6mm}
\end{figure}

To balance cost and robustness to cloneability (performance in the face of advanced fakes), this study endorses the use of synthetic physical CDP $\xhat$ \cite{belousov2022twins,belousov2024twins,pulfer2022wifs}, which is obtained through a trained deep mapper. This approach mitigates the high acquisition costs of $\xe$ while improving the low authentication accuracy associated with digital template $\t$. Synthetic $\xhat$ combines physical and digital benefits, providing a secure and cost-efficient solution. Building upon earlier works \cite{chaban2024eusipco, pulfer2022wifs}, our research primarily evaluates the effectiveness of synthetic CDP $\xhat$ against advanced machine-learning fakes $\f$ \cite{chaban2021fakes} through the scope of consumer mobile phones.

The primary focus is on a comparative analysis of the proposed authentication scenarios across different imaging systems. We conduct an analysis using both scanners and mobile phones to provide a comprehensive overview of various real-world conditions, utilizing a new dataset. Scanner, known for its high precision and consistency, provides reliable data for evaluation but is not considered as the practical use-case device for CDP verification. Thus, we test a range of mobile phones from older iPhone XS to cutting-edge iPhone 15 Pro Max. This diversity allows us to assess the authentication performance across multiple imaging systems, which vary significantly in terms of optical capabilities, sensor size, focal length, field of view, etc.. By including a broad spectrum of devices in our study, we aim to demonstrate that the authentication based on $\xhat$ can perform reliably under various conditions, while at the same given conditions the authentication based on $\t$ fails.

In our study, we pursue the following goals:
\begin{enumerate}
    \item to conduct the experiments using two industrial digital offset printers, which are a promising printing technology for CDP in view of the increasing demand for variable printing and product personalization; 
    \item to underline the vulnerabilities and limitations of nowadays authentication methodologies based on digital templates $\t$;
    \item to demonstrate the impact of different mobile phones on the authentication performance and on par with the flatbed scanner;
    \item to compare the existing authentication methods based either on digital template $\t$ or on physical template $\xe$ to the proposed one based on synthetic physical CDP $\xhat$;
    \item to outline the applicability scope of authentication based on $\xhat$.
\end{enumerate}

\begin{figure}
    \centering
    \begin{subfigure}{.49\textwidth}
        \includegraphics[width=\textwidth]{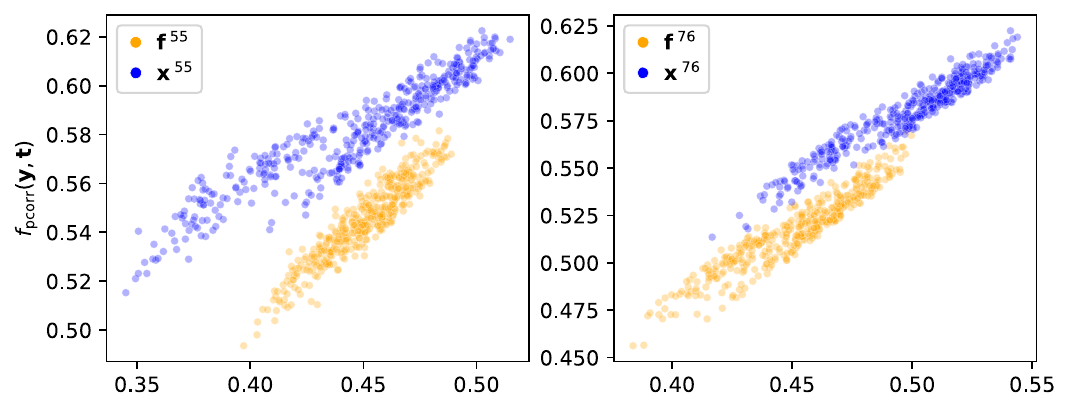}
        \caption{Old CDP samples from \cite{chaban2021fakes}.}
        \label{fig:scatter:scan:s1}
    \end{subfigure}
    \begin{subfigure}{.49\textwidth}
        \includegraphics[width=\textwidth]{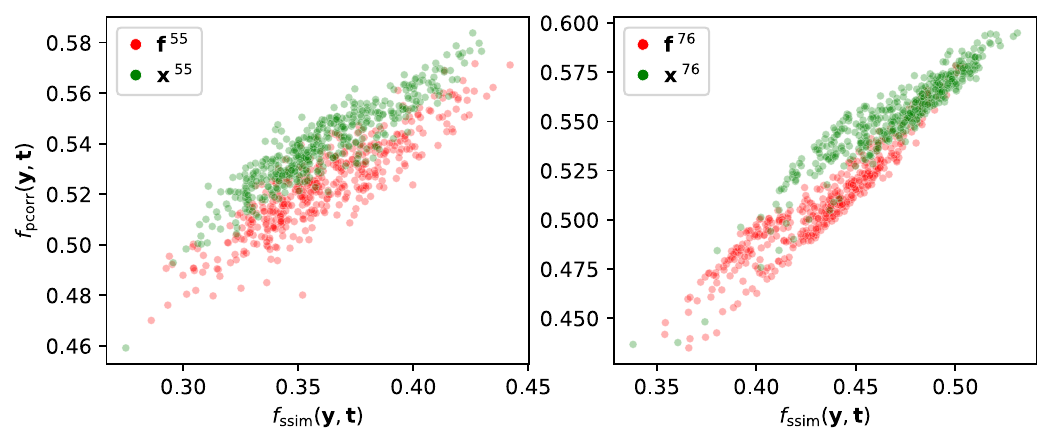}
        \caption{New samples created for this study.}
        \label{fig:scatter:scan:s2}
    \end{subfigure}
    \caption{The distributions of $\fpcorr(\y, \t)$ and $\fssim(\y, \t)$ for both printers, where probes $\bf y$ were scanned with a flatbed scanner. $\bf f$ and $\bf x$ denote fake and original probes, respectively, and $\bf t$ stands for the digital template.}
    \label{fig:scatter:scan}
    \vspace{-5mm}
\end{figure}

Meanwhile, we are aware that in such a study there is a multitude of possibilities and subjects for investigation, thus we have imposed the following limitations to deliver a clear message and for the sake of simplicity:

\begin{enumerate}
    \item We conduct experiments in an imaging-consistent manner, meaning that verification probe $\y$ is compared to reference $\r \in \{\xe, \xhat\}$ of the same imaging system.
    \item We consider the similarity assessment via two similarity metrics $\fsim$: Pearson Correlation Coefficient ($\fpcorr$) and Structural Similarity Measure Index\cite{wang2004ssim} ($\fssim$), since the cited studies have proven their effectiveness and to facilitate fair comparison and reproducibility.
    \item We choose iPhones since they provide reliable and consistent imaging across their device lineup, enabling a better track of imaging evolution.
\end{enumerate}

In conclusion, synthetic physical CDP $\xhat$ offers a promising solution to the vulnerabilities in traditional CDP systems by leveraging both physical and digital advantages for enhanced security and cost-effectiveness. Our study, involving scanner and mobile phone evaluations, aims to validate this solution and provide a robust framework for its implementation across various industries. The findings will contribute to the development of more secure anti-counterfeiting technologies, protecting critical products and documents.

%% file: sections/02_dataset.tex
\section{Dataset}

In this study, we introduce a novel dataset created using previously established methodologies and CDP design principles. The dataset was produced on two industrial digital offset printers: HP Indigo 5500 (HPI55) and 7600 (HPI76). Unlike previous studies \cite{chaban2021fakes,chaban2024eusipco,pulfer2022wifs}, the resulting difference between original and fake CDP is even less distinguishable both perceptually and statistically as demonstrated in Fig. \ref{fig:scatter:scan}. This in turn presents a more challenging problem for the classification \footnote{Dataset is available here: \url{https://github.com/romaroman/cdp-synthetics-dataset}.}.

The dataset consists of 144 unique CDP, each randomly generated with a binary distribution where 50\% of the pixels are black and 50\% are white. Such a decision was based on prior research indicating that this amount of CDP suffices to assess the general statistics of CDP. Consequently, the study involves 576 physical instances, derived from 2 printers × 2 origins (originals and fakes) × 144 unique CDP.

\begin{figure*}[!t]
    \centering
    \includegraphics[width=0.85\textwidth]{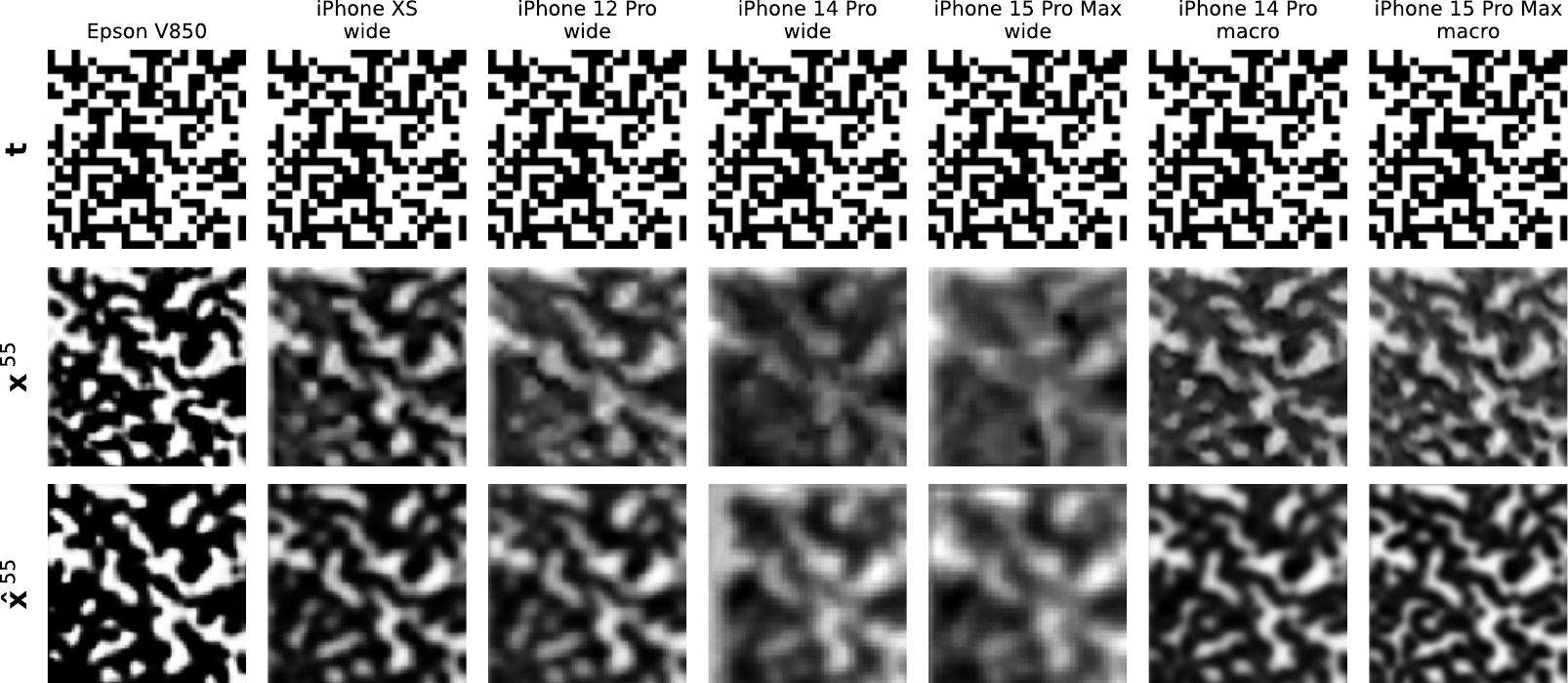}
    \caption{The examples of the left-upper crop of CDP \#166 printed on HPI55 and captured with a whole range of studied devices. Along with the digital template $\t$ and its printed form $\x$, the figure comprises the synthetic physical $\xhat$ that visually closely resembles real physical CDP $\x$.}
    \label{fig:examples}
    \vspace{-4mm}
\end{figure*}

\begin{figure*}[!t]
    \centering
    \includegraphics[width=0.9\textwidth]{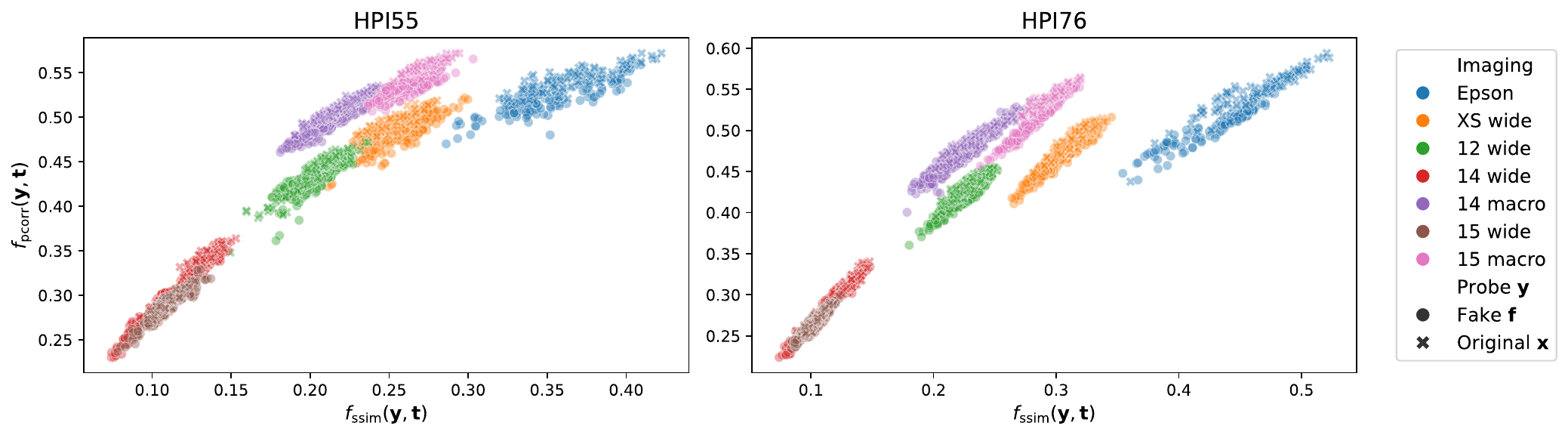}
    \caption{The scatter plots of both $\fsim$ for all given combinations of printer and imaging device.}
    \label{fig:scatter}
    \vspace{-6mm}
\end{figure*}

A key highlight of our new dataset collection is the number of used mobile phones, from the nearly obsolete iPhone XS (\textit{XS wide}) to the most advanced models currently available, such as the iPhone 15 Pro Max (\textit{15 wide}, \textit{15 macro}), which also includes the iPhone 12 Pro (12 wide) and iPhone 14 Pro (\textit{14 wide}, \textit{14 macro}). Notably, the iPhone 14 Pro and iPhone 15 Pro Max feature two relevant camera modules for this study: a regular wide and an ultra-wide macro lens. The macro lens, despite being a consumer-level component, already yields a much higher effective resolution, hypothetically surpassing the quality of Epson V850 (\textit{Epson}), the scanner used in our study. As shown in Fig. \ref{fig:scatter} the samples acquired with different imaging systems clearly possess distinguishable resulting statistics of similarity metric $\fsim(\y, \t)$, however, we can observe a certain overlap between \textit{14 wide} and \textit{15 wide} systems, and for \textit{14 macro} and \textit{15 macro}.

Each physical CDP, whether fake or original, was captured 3 times by every imaging device. After automatic quality control, a fraction of $\approx 5\%$ photos was discarded. This approach ensures high-quality images by excluding blurred or low-contrast captures. The effective PPI and a scale factor with respect to $\t$ differ between imaging systems. For authentication purposes, originals were divided into enrollment $\xe$ and verification probes $\y$.

The dataset was acquired using a custom-built mobile application, which enabled the capture of uncompressed images. This application was designed to obtain photos of the highest possible quality by:
\begin{enumerate}
    \item mitigating most causes of variability, such as illumination changes, motion blur, and out-of-focus issues;
    \item capturing from the minimal possible focal distance to yield the highest effective PPI;
    \item minimization of perspective transformation by guidance in the user interface.
\end{enumerate}

All obtained CDP were aligned with respect to the digital template using SIFT descriptors \cite{lowe2004distinctive}, followed by pixel-precision alignment using the same digital template.
 
In summary, our dataset, printed using HPI55 and HPI76 printers, and captured with six different mobile imaging systems, presents a more challenging authentication dataset by reducing the statistical distinguishability between original and fake CDP. This dataset, acquired by a number of modern consumer-grade mobile phones and a specialized application, will serve as a valuable resource for further studies on CDP-based authentication systems.

%% file: sections/03_methodology.tex
\section{Methodology}

In this section, we outline the methodology used in this study for the CDP authentication. The described steps are independently applied to each unique combination of industrial printer and imaging device.\footnote{GitHub repository: \url{https://github.com/romaroman/cdp-synthetics}.}

\subsection{General Overview}

\begin{figure*}
    \centering
    \begin{subfigure}{.9\textwidth}
        \includegraphics[width=\textwidth]{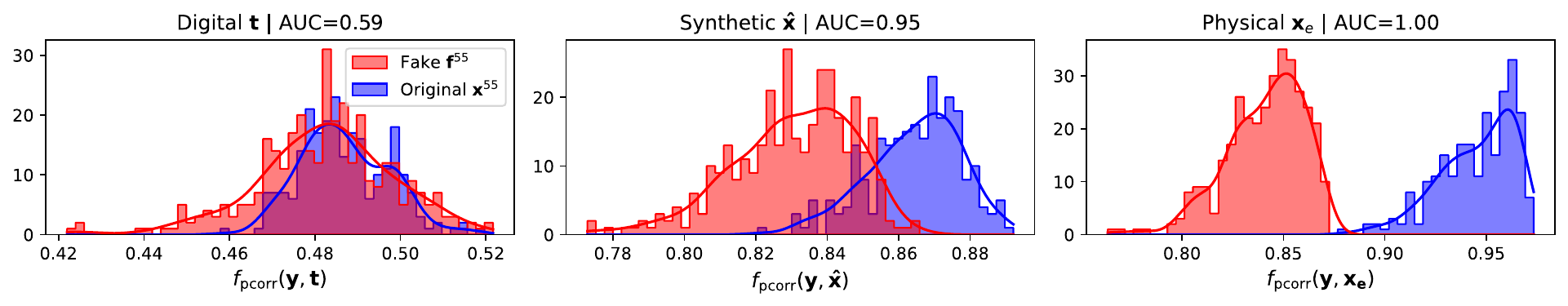}
        \caption{for HPI55}
        \label{fig:hist:55}
    \end{subfigure}
    \begin{subfigure}{.9\textwidth}
        \includegraphics[width=\textwidth]{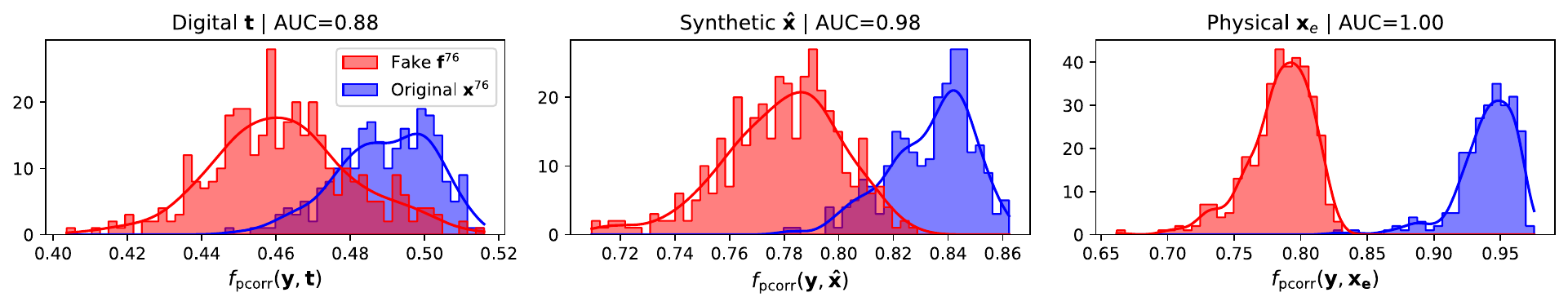}
        \caption{for HPI76}
        \label{fig:hist:76}
    \end{subfigure}
    \caption{The distributions of $\fpcorr(\y, \r)$ for CDP captured with the worst-case imaging system, \textit{iPhone XS wide}.}
    \label{fig:hist}
    \vspace{-5mm}
\end{figure*}

To create the original CDP $\x$, we begin by printing the digital template $\t$. To generate the fake CDP $\f$, we use the original CDP $\x$ as input to a pre-trained U-Net \cite{chaban2021fakes,ronneberger2015unet} estimator $A$, which produces an estimated digital template $\hat{\t}$ as shown by the red dashed line in Fig. \ref{fig:lifecycle}. This estimated template $\hat{\t}$ is then reprinted using the same printer and settings, resulting in the fake CDP $\f$. The complete description of fakes $\f$ production details can be found in \cite{chaban2021fakes}.

The original CDP $\x$ is divided into two subsets. The first subset is used both for enrollment denoted as $\xe$ and for training the pix2pix model \cite{isola2017pix2pix} to generate synthetic physical CDP $\xhat$. Effectively, for the training of synthetic generator $G$ only approximately 30 CDPs are sufficient, and such generator is well-scalable for much bigger volumes \cite{belousov2024twins}.

The second subset of $\x$ is reserved for use as incoming probes $\y$ during authentication. These probes $\y$ (which can be either original $\x$ or fake $\f$) are compared against references $\r$, (which can be digital template $\t$, synthetic physical $\xhat$, or enrolled $\xe$ CDPs) to calculate the similarity score $\fsim(\y, \r)$.

\subsection{Training of Synthetic CDP Generator}

The pix2pix model \cite{isola2017pix2pix} $G$ is used to learn the mapping from the digital template CDP $\t$, to the synthetic physical CDP $\xhat$. We employed a state-of-the-art paired pix2pix architecture, where a U-Net \cite{ronneberger2015unet} serves as the generator $G$ and PatchGAN is used as the discriminator $D$. The training process involves optimizing specific loss functions for both the discriminator and the generator to achieve high-quality synthetic CDP for pairs ${\{\t_i, \x_i\}}^N_{i=1}$, where $N = 40$ is the number of training pairs.

The discriminator $D$ is trained to distinguish between real physical CDP $\xe$ and synthetic physical CDP $\xhat$ generated by the generator $G$. Its loss function is defined as:

\begin{align}
    \loss_D &= - \mathbb{E}_{(\t, \x)} \left[ \log D(\t, \x) \right] - \mathbb{E}_{\t, \xhat = G(\t)} \left[ \log (1 - D(\t, \xhat)) \right],
\end{align}
where $\mathbb{E}_{(\t, \x)}\left[\bullet\right]$ denotes the mathematical expectation.

This loss function encourages $D$ to assign high probabilities to real CDP and low probabilities to synthetic ones, thereby becoming sensitive to identifying real CDP.

Conversely, $G$ is trained to produce synthetic physical $\xhat$ that are indistinguishable from $\x$. Its loss function is given by:

\begin{align}
    \loss_G &= - \mathbb{E}_{\t, \xhat = G(\t)} \left[ \log D(\t, \xhat) \right] + \lambda \mathbb{E}_{(\t, \x)} \left[ \| \x - G(\t) \|_1 \right],
\label{eq:loss} 
\end{align}
where $\lambda$ controls the trade-off between two terms.

This loss function encourages $G$ to generate samples that $D$ misclassifies as real, while also maintaining a high degree of visual and structural fidelity to $\x$ by including a similarity term weighted by $\lambda$.

By balancing these loss functions, the pix2pix model effectively learns to generate $\xhat$ that are both statistically and visually similar to $\x$. The generator's output is thus guided to not only deceive the discriminator but also to closely resemble real CDP, ensuring that the synthetic CDP can be used reliably in place of original CDP for authentication purposes.

For model training we have conducted a train/test split by ensuring that the model trains on a small fraction ($\approx30\%$) of $\xe$ and is tested on the previously unseen $\t$, effectively generating $\xhat$ which were never subjected to loss computation and model optimization.

This approach ensures that the pix2pix model effectively learns to generate high-quality synthetic CDP that are not only visually indistinguishable from real CDP but also maintain a high degree of similarity from a statistical point of view.

\subsection{Similarity Assessment}

In the authentication stage, we compute the similarity metrics $\fpcorr$ and $\fssim$, for the incoming probe $\y \in \{\x, \f\}$ with respect to the reference $\r \in \{\t, \xhat, \xe\}$. These similarity scores are then analyzed using Receiver Operating Characteristic (ROC) curves, and the corresponding Area Under the Curve (AUC) values are calculated to evaluate the effectiveness of each reference $\r$. We do not create a specific threshold-based classifier, as the ROC analysis provides a sufficient measure of potential classification performance in our study. Obviously, in practical use cases, an appropriate decision threshold should be chosen based on the provided ROC curves and the desired probability of false acceptance and probability of miss.

%% file: sections/04_results.tex
\section{Results}

\begin{figure*}
    \centering
    \begin{subfigure}{.47\textwidth}
        \includegraphics[width=\textwidth]{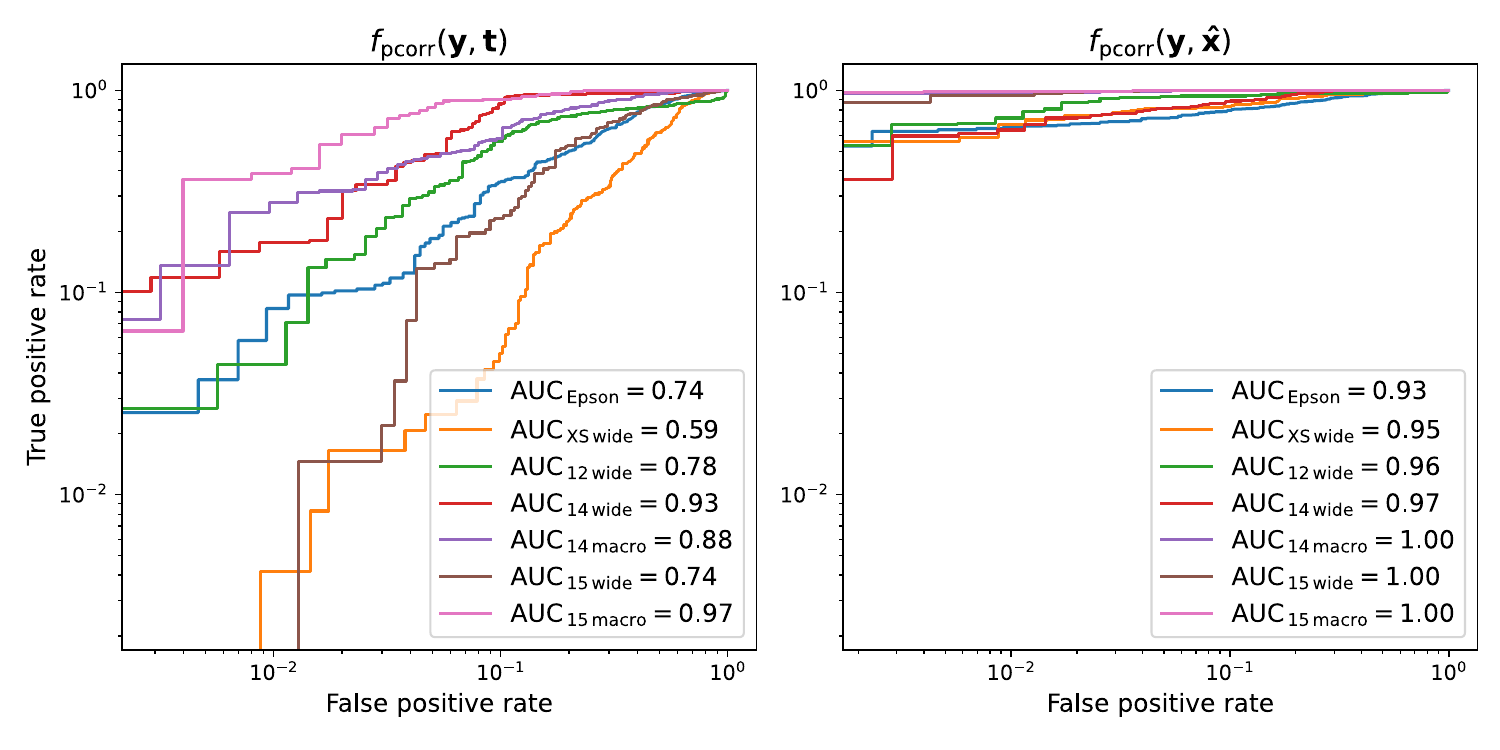}
        \caption{ROC curves for HPI55}
        \label{fig:roc:55}
    \end{subfigure}
    \begin{subfigure}{.47\textwidth}
        \includegraphics[width=\textwidth]{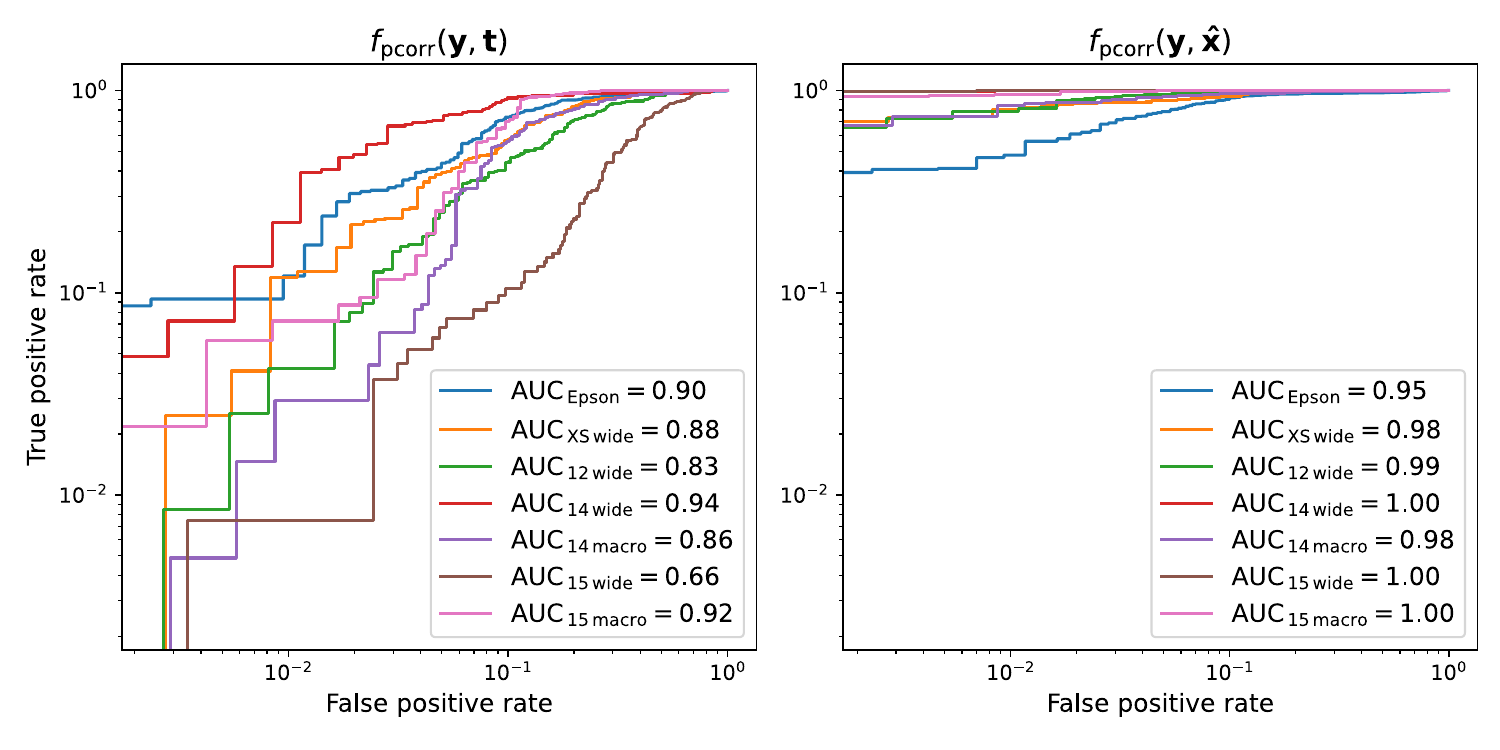}
        \caption{ROC curves for HPI76}
        \label{fig:roc:76}
    \end{subfigure}
    \caption{ROC curves for $\fpcorr(\y, \r)$, where $\r \in \{\t, \xhat\}$, with corresponding AUC values for each imaging system. $\xe$ is not shown as its AUC equals 1.00 for all systems. $\xhat$ were trained for each imaging system.}
    \label{fig:roc}
\end{figure*}

\input{sections/tables/aucs}

Referring back to Fig. \ref{fig:scatter}, we conducted same-device experiments, where the incoming probe $\y$ is compared to $\xe$ or $\xhat$ from the same physical and acquisitional origin. Future work should explore cross-device testing, as some imaging systems (like \textit{14 macro} and \textit{15 macro}) that share similar statistical distributions as claimed by the vendor.

As expected from the distributions shown in Fig. \ref{fig:scatter:scan}, we observed similar degrees of separability in terms of $\fsim(\y, \t)$ for mobile phones. This is well-supported in Fig. \ref{fig:hist} (Digital $\t$ plots), and the trend is consistent for both printers. The AUC values indicated in the plots' titles correspond to the ROC curves in Fig. \ref{fig:roc}. Table \ref{tab:aucs} presents a comprehensive comparison of $\fsim(\y, \t)$ performance across all imaging systems. Key conclusions for $\fsim(\y, \t)$ are:

\begin{enumerate}
    \item Authentication accuracy increases with imaging effective resolution (from old devices to more recent ones).
    \item Old phones demonstrate unsatisfactory performance for the authentication based on digital template $\t$.
    \item Phones equipped with macro lenses sometimes outperform scanners, approaching ideal performance.
    \item Mobile phone sampling shows high variability and uncertainty, causing considerable fluctuations in AUC.
    \item On average, $\fssim$ is less discriminative than $\fpcorr$.
    \item Neither printer nor imaging system combination provides satisfactory AUC.
\end{enumerate}

Next, we assessed the performance of $\fsim(\y, \xe)$ as the physical baseline reference. As shown in Fig. \ref{fig:hist} (Physical $\xe$ plots), even the oldest mobile phone shows negligible overlap between $\x$ and $\f$ for both printers, underscoring the superiority of $\fsim(\y, \xe)$. This trend is consistent across all acquisitions. When $\y = \x$, we effectively compute $\fsim(\x, \xe)$, where both $\x$ and $\xe$ originate from the same physical CDP, achieving the highest mutual information. Consequently, $\fsim(\f, \xe)$ cannot match this level of microstructural similarity. Table \ref{tab:aucs} shows perfect separability for any printer and imaging device pair, except for \textit{12 wide} due to alignment issues discovered through manual investigation. Therefore, ROC curves for these results are redundant and not shown in this paper. Key findings can summarized as:

\begin{enumerate}
    \item Results align well with previous studies\cite{chaban2024eusipco} and are explained by $\xe$ being a physically unclonable function \cite{maes2013puf} of CDP.
    \item $\fsim(\y, \xe)$ represents a sufficient statistics for authentication systems.
    \item Authentication based on physical CDP $\xe$ achieves excellent results even for \textit{XS wide}. However, such an approach requires enrollment from each physical CDP $\x$.
\end{enumerate}

With these baselines, we assess the viability of synthetic physical $\xhat$. No enrollment from all future physical objects is necessary, once the generator $G$ is trained on a small fraction of enrolled physical $\xe$. Fig. \ref{fig:hist} (Synthetic $\xhat$ plots) shows significant improvement in AUC for $\fpcorr(\y, \xhat)$ compared to $\fpcorr(\y, \t)$, regardless of the printer. This trend is consistent across all imaging devices, as shown in Table \ref{tab:aucs} and Fig. \ref{fig:roc}. In many scenarios, $\xhat$ achieves perfect separability (AUC = 1), though this might be due to a small sample size and may decrease with a larger number of samples. Key takeaways are:

\begin{enumerate}
    \item $\xhat$ as reference $\r$ significantly increases separability of $\x$ from $\f$ for low quality imaging devices.
    \item For scenarios where AUC $<1$, improvements such as fine-tuning, data refinement, and tweaking network architecture or loss functions are still possible.
    \item Although $\xhat$ lags behind $\xe$ in terms of AUC, it is sufficient for certain imaging systems in its current implementation.
    \item AUC for scanner-based results is the lowest compared to any mobile phone, suggesting that the synthetic generator $G$ is more suited for the mobile phone images.
    \item Slightly better performance of $\fssim$ for HPI55 with \textit{12 wide} pair may be linked to the generator $G$'s loss function, which includes SSIM, L2, and L1 norms.
\end{enumerate}

In summary, the results highlight the separability of original $\x$ and fake $\f$ samples across various imaging and printing systems. $\fpcorr(\y, \xhat)$ yields higher AUC values compared to $\fsim(\y, \t)$. Mobile phones show high variability in AUC, with macro lenses performing close to ideal in some cases. The $\fsim(\y, \xe)$ approach consistently achieves perfect separability due to the physically unclonable nature of $\xe$. The synthetic reference $\xhat$ significantly enhances separability, though still below $\xe$’s performance. Scanner-based results exhibit the lowest AUC, indicating that the generator $G$ adapts better to mobile phone conditions. Overall, these findings underscore the potential for improved authentication using $\xe$ and $\xhat$ references, with implications for low-end device applications. The obtained results are based on the pix2pix model, which is a component of the more powerful TURBO model \cite{quetant2022turbo}, known for its superior performance in CDP applications \cite{belousov2024twins}. Therefore, we anticipate even better results with an increase in model complexity.

%% file: sections/tables/aucs.tex
\begin{table*}[htbp]
\centering

\renewcommand{\arraystretch}{1.3}

\begin{tabular}{ll|ccccccc|ccccccc}
\toprule
 &  & \multicolumn{7}{c|}{HPI55} & \multicolumn{7}{c}{HPI76} \\

 $\r$ & $\fsim$ & \rotatebox[origin=c]{90}{\textit{Epson}} & \rotatebox[origin=c]{90}{\textit{XS wide}} & \rotatebox[origin=c]{90}{\textit{12 wide}} & \rotatebox[origin=c]{90}{\textit{14 wide}} & \rotatebox[origin=c]{90}{\textit{15 wide}} & \rotatebox[origin=c]{90}{\textit{14 macr}o} & \rotatebox[origin=c]{90}{\textit{15 macro}} & \rotatebox[origin=c]{90}{\textit{Epson}} & \rotatebox[origin=c]{90}{\textit{XS wide}} & \rotatebox[origin=c]{90}{\textit{12 wide}} & \rotatebox[origin=c]{90}{\textit{14 wide}} & \rotatebox[origin=c]{90}{\textit{15 wide}} & \rotatebox[origin=c]{90}{\textit{14 macro}} & \rotatebox[origin=c]{90}{\textit{15 macro}} \\
\midrule
\multirow[t]{3}{*}[-0.75em]{$\t$} & $\fpcorr$ & 0.74 & 0.59 & 0.78 & 0.93 & 0.74 & 0.88 & 0.97 & 0.90 & 0.88 & 0.83 & 0.94 & 0.66 & 0.86 & 0.92 \\
                                & $\fssim$ & 0.51 & 0.71 & 0.68 & 0.90 & 0.64 & 0.74 & 0.79 & 0.81 & 0.78 & 0.69 & 0.93 & 0.63 & 0.81 & 0.81 \\
\cline{1-16}
\multirow[t]{3}{*}[-0.75em]{$\xhat$} & $\fpcorr$ & 0.93 & 0.95 & 0.96 & 0.97 & 1.00 & 1.00 & 1.00 & 0.95 & 0.98 & 0.99 & 1.00 & 1.00 & 0.98 & 1.00 \\
                                   & $\fssim$ & 0.93 & 0.96 & 0.97 & 0.96 & 1.00 & 0.99 & 1.00 & 0.94 & 0.98 & 0.99 & 1.00 & 1.00 & 0.97 & 1.00 \\
\cline{1-16}
\multirow[t]{3}{*}[-0.75em]{$\xe$} & $\fpcorr$ & 1.00 & 1.00 & 0.97 & 1.00 & 1.00 & 1.00 & 1.00 & 1.00 & 1.00 & 1.00 & 1.00 & 1.00 & 1.00 & 1.00 \\
                                 & $\fssim$ & 1.00 & 1.00 & 0.99 & 1.00 & 1.00 & 1.00 & 1.00 & 1.00 & 1.00 & 1.00 & 1.00 & 1.00 & 1.00 & 1.00 \\

\bottomrule
\end{tabular}
\caption{The breakdown of AUC values presented for the whole range of similarity metrics, references, imaging systems and printers of current study scope.}
\label{tab:aucs}
\vspace{-6mm}
\end{table*}

%% file: sections/05_conclusions.tex
\section{Conclusions and future work}

In this study, we have explored the potential of synthetic physical $\xhat$ as a viable alternative to the existing authentication framework. Our research demonstrates that synthetic physical $\xhat$ significantly improves authentication accuracy compared to digital templates $\t$, with consistent performance across various imaging systems.

Our findings indicate that synthetic physical $\xhat$ can substantially enhance the separability between original $\x$ and fake $\f$ samples. While mobile phones exhibit some variability in authentication performance, the use of ultra-wide macro lenses pushes the frontier of mobile phone-based authentication of CDP by showing near-ideal results in certain cases. Although $\xhat$ does not yet match the perfect performance of physical enrollment $\xe$, it demonstrates significant potential. Further refinements in model training, data augmentation, and network architecture could enhance its performance even further.

The study supports the adoption of synthetic $\xhat$ over digital templates $\t$ or physical enrollment $\xe$ in scenarios where cost-effectiveness and scalability are critical. This research validates that $\xhat$ can perform reliably under various conditions, offering a robust framework for real-world anti-counterfeiting applications across different industries.

Future work will include comprehensive cross-device testing to investigate the robustness of synthetic $\xhat$. This will involve examining how estimators trained on one mobile phone model perform with probes from different systems. We will also explore the integration of other similarity metrics and advanced machine-learning techniques to enhance the accuracy and reliability of CDP-based authentication. Expanding the dataset and incorporating more diverse imaging systems will provide further insights into the practical applications of synthetic CDP.

Overall, this research contributes to the development of more secure and economical anti-counterfeiting technologies, providing a valuable resource for the protection of critical products. The use of old mobile phones might be of interest to developing countries, where the fraction of fakes is relatively high and the mobile phone market is dominated by older models.